\documentclass[trackchanges, twocolumn]{aastex7}
\usepackage{chemmacros}

\begin{document}

\title{The He I D3 Line as a Proxy for Magnetic Activity using EXPRES Solar Observations}

\author[orcid=0009-0007-0922-7315,sname='Ellwarth']{Momo Ellwarth}
\affiliation{Lowell Observatory, 1400 West Mars Hill Road, Flagstaff, AZ 86001, USA}
\affiliation{Department of Astronomy and Planetary Science, Northern Arizona University, PO Box 6010, Flagstaff, AZ 86011 USA}
\email[show]{mellwarth@lowell.edu}

\author[orcid=0000-0003-4450-0368,sname='LLama']{Joe Llama}
\affiliation{Lowell Observatory, 1400 West Mars Hill Road, Flagstaff, AZ 86001, USA}
\email[]{jllama@lowell.edu}

\author[orcid=0000-0002-3852-3590]{Lily L.\ Zhao}
\thanks{NASA Sagan Fellow}
\affil{Department of Astronomy \& Astrophysics, University of Chicago, Chicago, IL, USA}
\email[]{lilylingzhao@uchicago.edu}

\author[orcid=0000-0002-9873-1471,sname='Brewer']{John M. Brewer}
\affiliation{Department of Physics and Astronomy, San Francisco State University, 1600 Holloway Avenue, San Francisco, CA 94132,, USA}
\email[]{jmbrewer@sfsu.edu}
\author[orcid=0000-0002-4974-687X,sname='Szymkowiak']{Andrew Szymkowiak}
\affiliation{Department of Astronomy, Yale University, 52 Hillhouse Avenue, New Haven, CT 06511, USA}
\email[]{andrew.szymkowiak@yale.edu}

\author[orcid=0000-0002-9288-3482,sname='Roettenbacher']{Rachael M. Roettenbacher}
\affiliation{Department of Astronomy, University of Michigan, 1085 S. University Ave., Ann Arbor, MI 48109, USA}
\email[]{rmroett@umich.edu}
\author[orcid=0000-0002-3522-5846,sname='Sikora']{James T. Sikora}
\affiliation{Lowell Observatory, 1400 West Mars Hill Road, Flagstaff, AZ 86001, USA}
\email[]{jsikora@lowell.edu}
\author[orcid=0000-0001-7047-8681,sname='Polanski']{Alex S. Polanski}
\affiliation{Lowell Observatory, 1400 West Mars Hill Road, Flagstaff, AZ 86001, USA}
\email[]{apolanski@lowell.edu}

\author[orcid=0000-0001-7032-8480,sname='Saar']{Steven H. Saar}
\affiliation{Center for Astrophysics | Harvard \& Smithsonian, 60 Garden St., Cambridge, 02138 MA USA}
\email[]{ssaar@cfa.harvard.edu}

\begin{abstract}

Stellar activity remains one of the primary challenges in the detection and characterization of low-mass exoplanets, as it can induce radial velocity (RV) variations that mask or mimic planetary signals. Identifying reliable activity proxies is essential in order to distinguish stellar variability from genuine planetary signatures. In this study, we examine the variability of the chromospheric He I D3 line in high-resolution solar spectra and assess its potential as an activity indicator. We find a strong correlation between the He I D3 line intensity variation and the Sun’s unsigned magnetic flux derived from SDO/HMI data as well as with the solar RVs. Our results suggest that the He I D3 line offers a promising and straightforward proxy for magnetic activity, which may complement existing stellar activity indicators. Its inclusion could help disentangle stellar signals in RV measurements and ultimately improve the detection of Earth-like exoplanets.

\end{abstract}

\keywords{\uat{Radial velocity}{1332}; \uat{Solar magnetic fields}{1503}; \uat{The Sun}{1693}; \uat{Stellar activity}{1580}; \uat{Solar activity}{1475}}

\section{Introduction} 

Detecting small, Earth-like planets around Sun-like stars remains a major challenge in modern exoplanet research due to the effects of stellar activity itself \citep[][]{2021arXiv210714291C}. The radial velocity (RV) method is one of the most widely used techniques to detect low-mass planets. Planetary masses are commonly estimated with the RV method and are also a key parameter for the characterization of exoplanet atmospheres \citep{2019ApJ...885L..25B}.
However, in cool stars such as the Sun, stellar variability can both obscure and mimic planetary signals, thereby complicating the interpretation of RV data \citep[e.g.][]{1997ApJ...485..319S, 2011A&A...527A..82D}.
Solar‑like p‑modes produce RV variations of (0.1–3.0)\,m\,s$^{-1}$ \citep[][]{2000ssma.book.....S}, with granulation introducing a comparable magnitude of variance \citep[e.g.][]{2015A&A...583A.118M}. Whereas active regions such as spots, faculae, and plage can drive amplitudes up to 10\,m\,s$^{-1}$ \citep[e.g.][]{2010A&A...512A..39M}.
In contrast, an Earth analogue around a solar twin induces only a semi‑amplitude of 9\,cm\,s$^{-1}$. Consequently, to reach Earth-like sensitivity, robust stellar activity proxies and forward modeling are required to disentangle activity‑driven variability from a planet’s barycentric reflex motion.

While long-baseline optical interferometry has enabled spatially resolved imaging of stars and even their spots \citep[e.g.][]{2007Sci...317..342M,2016Natur.533..217R}, the Sun remains the only star for which we can obtain continuous, high-resolution, disk-resolved time series measurements. This makes the Sun an unparalleled benchmark for understanding stellar variability and the resulting intrinsic RV variations.

During the past decades, the Sun has been continuously monitored by several high-precision spectrographs in a Sun-as-a-star mode, including HARPS-N \citep{2012SPIE.8446E..1VC} with its solar feed \citep{2015ApJ...814L..21D}, NEID (\cite{2016SPIE.9908E..7HS} which gained a solar-feed in 2020 \citep{2022AJ....163..184L}), EXPRES \citep{2016SPIE.9908E..6TJ} in combination with the Lowell Observatory Solar Telescope \citep[LOST;][]{2022BAAS...54e.102L}, and more recently the Keck Planet Finder (KPF; \cite{2016SPIE.9908E..70G}) fed by the Solar Calibrator (SoCal; \citealt{2023PASP..135l5002R}). In addition to this wealth of disk-integrated observations, dedicated space missions provide complementary disk-resolved measurements. A prime example is NASA’s Solar Dynamics Observatory (SDO; \citealt{2012SoPh..275....3P}), which has been observing the Sun since 2010. SDO's Helioseismic and Magnetic Imager (HMI; \citealt{2012SoPh..275..207S}) continuously maps the Sun’s magnetic field at high spatial and temporal resolution. By capturing the full solar disk, HMI provides detailed insights into the distribution and evolution of magnetic activity across the surface, offering a unique complement to Sun-as-a-star observations. Additionally, long-term synoptic networks and space missions, including BiSON \citep{1996SoPh..168....1C}, GONG \citep{1996Sci...272.1284H}, SOHO \citep{2012SoPh..275....3P}, SOLIS \citep{1998ASPC..154..636K}, and SORCE \citep{2000SPIE.4135..192W}, have provided continuous coverage over multiple decades
covering between 2 and 4 full solar activity cycles. These rich, multi-decade data sets allow us to study activity-driven changes in unparalleled detail, enabling insights into processes that no other star can currently offer.

\cite{2016MNRAS.457.3637H} used SDO/HMI solar observations to introduce the unsigned magnetic flux $|B_{obs}|$ of the Sun as a proxy for RV variations of Sun-like stars. In \cite{2022ApJ...935....6H} they found a correlation between $|B_{obs}|$ and the solar RV with a Spearman correlation coefficient of 0.92. The authors used this magnetic flux to model the corresponding solar RVs but still found RV residuals (between model and observed RVs) with an rms of 0.56\,m\,s$^{-1}$. During the quiet phase of the Sun the rms even increased to 1.02\,m\,s$^{-1}$. They still found a rotationally modulated variation in their residuals, particularly during times of high magnetic activity. This shows that we still need to work on our understanding of stellar variability in order to successfully detect Earth-like planets using RV measurements.

One established chromospheric activity indicator is the \ch{He I} D3 triplet (hereafter referred to as \ch{He} D3) at the vacuum wavelength $\lambda \approx $ 5877.24\,\AA\, \footnote{Kramida, A., Ralchenko, Yu., Reader, J. and NIST ASD Team (2024). NIST Atomic Spectra Database (version 5.12), [Online]. Available: https://physics.nist.gov/asd [Fri Oct 24 2025]. National Institute of Standards and Technology, Gaithersburg, MD. DOI: https://doi.org/10.18434/T4W30F}. This triplet consists of six possible transitions, with the strongest component at the mentioned wavelength. \cite{1997A&A...326..741S} found good correlation between He D3 absorption line and the emission of the Ca II H and K lines for G and K dwarfs. He D3 absorption is typically absent in quiet, non-magnetic solar regions \citep[][]{1981ApJ...244..345L}, indicating that, unlike other activity-sensitive lines, it is not formed through acoustic heating. This makes He D3 a particularly clean tracer of magnetic activity as it becomes visible only in the presence of magnetically active regions. \citet{2008ApJ...676..628S} investigated star–planet interactions of F, G, and K stars by monitoring chromospheric activity indicators, including the He D3 line. They measured the equivalent width of the He D3 feature in a sample of 8 planet-hosting stars. While they were unable to detect clear activity-induced variations in the He D3 line, they found an obvious increase in the line's absorption for two of their stars coinciding with a maximum emission of Ca II. Their results suggest that the line has potential as an activity proxy and further investigation is warranted.

Although the \ch{He} D3 line has been identified and used in several early studies \citep[e.g.][]{1981ApJ...244..345L,1984BAAS...16..940V,1997A&A...326..741S}, it has seen relatively little application in practice for solar-type stars, especially compared to well-established proxies like Ca II H\&K or H$\alpha$. One reason is the fact that the already weak line is blended with tellurics as well as an unidentified line (blended with \ch{H2O}) at 5877.34\,\AA. 
These obstacles are likely the main reasons why it is difficult to measure the He D3 absorption or equivalent width in a consistent way.

In contrast, for M dwarfs, the \ch{He} D3 line has received somewhat more attention in recent years, with several studies exploring its potential as a chromospheric activity tracer in this stellar class. Generally, M‑dwarfs exhibit the \ch{He} D3 line in emission \citep[e.g.][]{2019A&A...623A..44S, 2022MNRAS.512.5067K}, and is known to correlate well with \ch{H $\alpha$} emission \citep{2002AJ....123.3356G, 2019A&A...623A..44S}, which suggests that these lines form in the same chromospheric region. However, in \cite{2009MNRAS.400..238H} they investigated 37 M dwarfs but only detected the He D3 line for high activity level stars and two intermediate active stars. In a year‑long HARPS campaign on the young M dwarf AU Mic, \citet{2022MNRAS.512.5067K} found that the He D3 index is strongly modulated at the stellar rotation period, with a best‑fit value of 4.79\,$\pm$\,0.01\,d, consistent with the rotation seen in other chromospheric tracers. Although the He D3 line clearly traces the stellar rotation, the authors found that it does not correlate with the RV variations (Pearson r = -0.145).

In this work, we reassess the diagnostic power of He D3 using high-resolution, full-disk solar observations from the EXtreme PREcision Spectrograph \citep[EXPRES;][]{2016SPIE.9908E..6TJ}, and explore its utility as a stellar activity proxy for Sun-like stars.

\begin{figure}
    \centering
    \includegraphics[width=\linewidth]{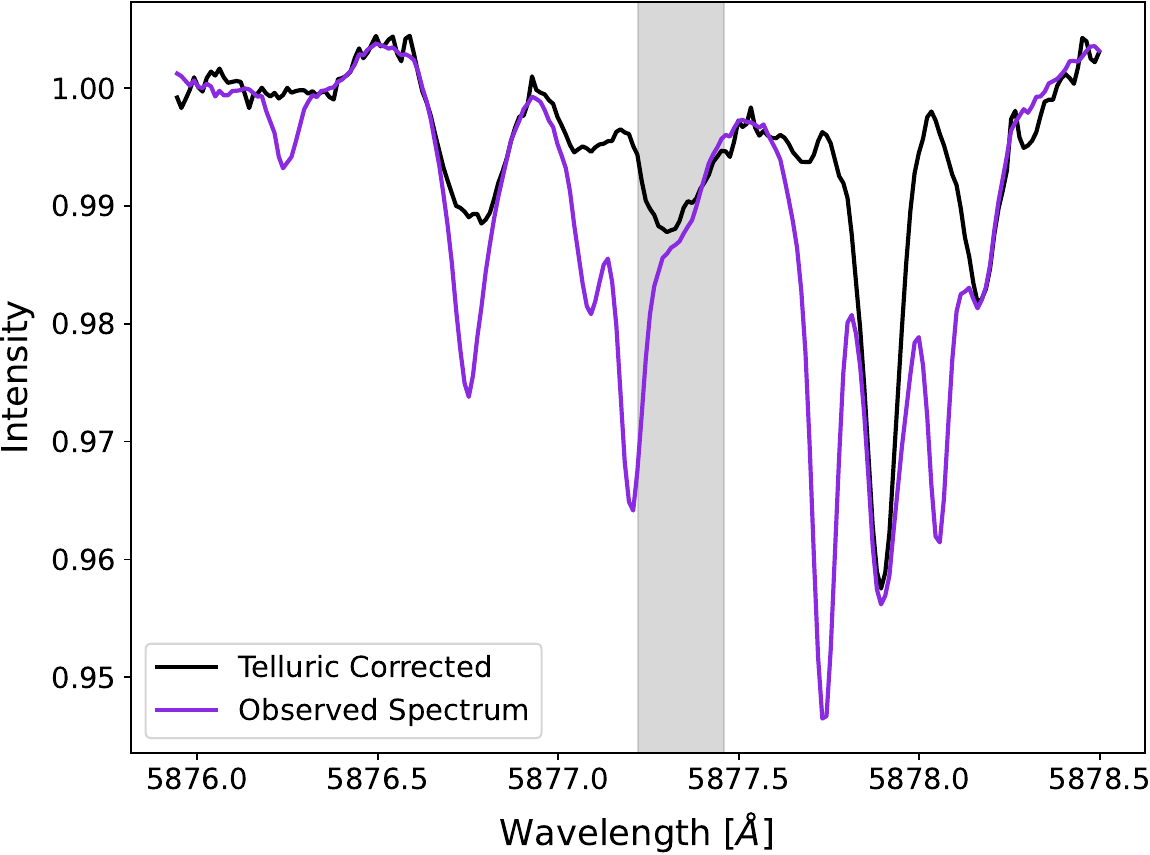}
    \caption{Averaged EXPRES spectrum over the year 2024. Purple: Observed Spectrum with telluric lines. Black: Telluric corrected spectrum. The gray area shows the wavelength range of interest.}
    \label{fig:HeD3_spec}
\end{figure}

\section{Data}
\subsection{EXPRES Solar Data} \label{sec:Data}

For our analysis, we used solar observations taken with LOST \citep{2022BAAS...54e.102L}, which feeds into EXPRES; installed on the 4.3-m Lowell Discovery Telescope (LDT) \citep{2012SPIE.8444E..19L} in Northern Arizona. LOST is a 70\,mm lens solar telescope, which is connected to an integrating sphere to obtain Sun-as-a-star solar spectra. EXPRES operates at high resolution (R = 137,500) and observes in the optical range between $\lambda =$ (4000-8000)\,\AA. The Sun-as-a-star exposures use a dynamic exposure meter built into EXPRES to terminate exposures once they reach a signal-to-noise ratio (SNR) of 500. In clear conditions, this typically results in an exposure time of about 180\,s. The LOST/EXPRES data are reduced using the standard EXPRES pipeline \citep{petersburg2020} with minimal alterations to account for the difference in barycentric correction. 

For our investigation of the He D3 triplet (vacuum wavelength $\lambda \approx$ 5877.25\,\AA\,), we analyzed EXPRES solar observations obtained between October 2020 and March 2025 (899 days of observations in total). This time span covers the transition from the minimum of the 24th solar cycle to the current maximum in Solar Cycle 25, providing a valuable dataset across a wide range of solar activity levels. Since the He D3 line is partially blended with Earth's telluric absorption features, we use the self-calibrated \texttt{SELENITE} telluric model that forms part of the standard EXPRES data reduction suite \citep{leet2019}. An example telluric-corrected spectrum is shown in Fig.~\ref{fig:HeD3_spec}.

\subsection{Unsigned Magnetic Flux} \label{sec:SDO}

The unsigned magnetic flux of a star is a widely used proxy for activity-related RV variations, as it reflects the total magnetic energy on the stellar surface. These activity induced RV variations are often rotationally modulated, following the stellar rotation period \citep[see, e.g.,][]{2022ApJ...935....6H}. In this study, we use the Sun’s unsigned magnetic flux to test whether the He D3 line traces similar activity-driven changes.

To obtain the unsigned magnetic flux values of the Sun, we use data collected by NASA's SDO, which has been continuously monitoring the Sun since 2010. SDO enables consistent and high-cadence tracking of the Sun’s surface magnetic activity over multiple solar cycles. In particular, the Helioseismic and Magnetic Imager (SDO/HMI) is designed to study solar oscillations and the surface magnetic field of the Sun. SDO/HMI provides full-disk continuum images centered around Fe\,I\,(6173\,\AA), dopplergrams, and both line-of-sight and vector magnetograms. HMI provides disk-resolved images of the solar disk with a resolution of 1'' with a 45\,s cadence. For this study, we downloaded the intensitygram, dopplergram, and magnetogram taken at 20:00 UTC (corresponding to local noon in Arizona) for each day that we have LOST/EXPRES observations. The data were then analyzed using the publicly available \texttt{SolAster} pipeline \citep{2022AJ....163..272E}, which provides time series of various solar parameters, including radial velocities, filling factors, and magnetic fluxes.

\begin{figure}
    \centering
    \includegraphics[width=\linewidth]{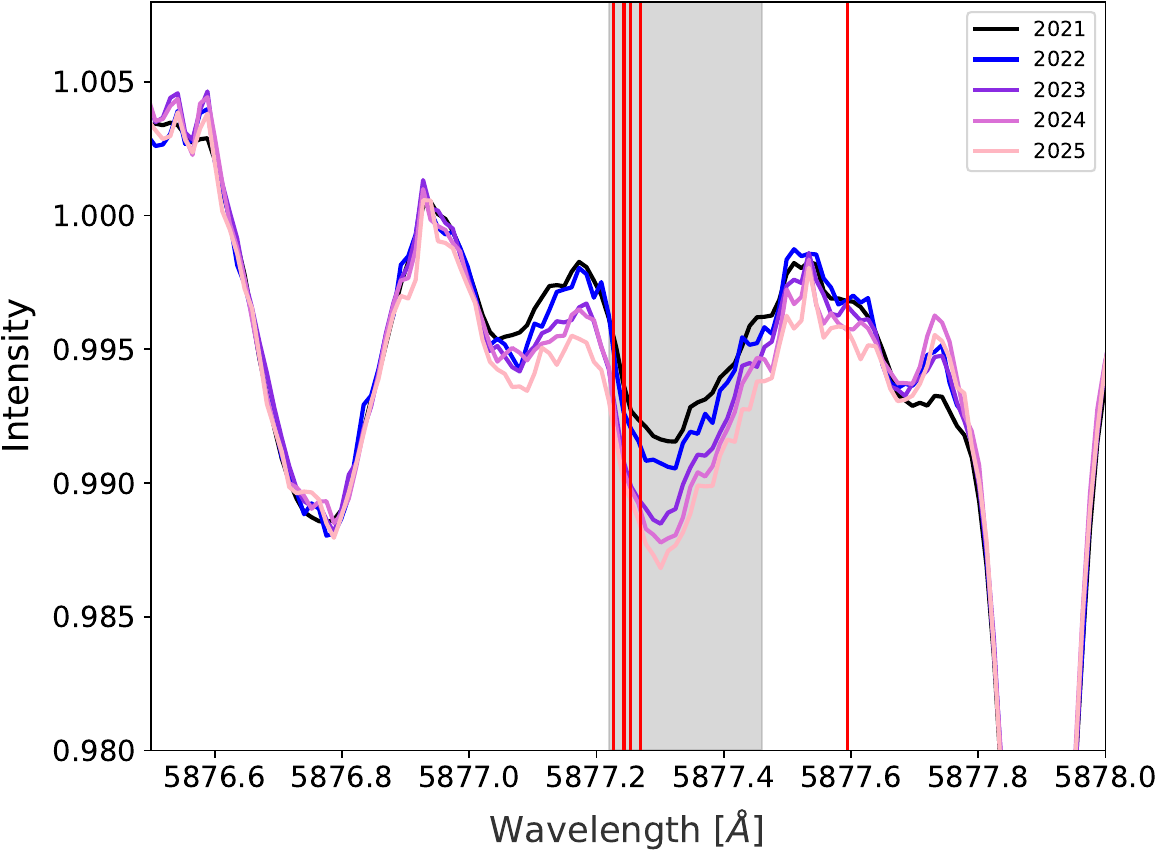}
    \caption{EXPRES spectra of the He I D3 line, averaged by year from 2021 to 2025. The red lines mark the laboratory wavelengths of the He D3 line (Table \ref{tab:table0}). The gray area mark the wavelength range used to calculate the line index.}
    \label{fig:HeD3_years}
\end{figure}

\begin{figure}
    \centering
    \includegraphics[width=\linewidth]{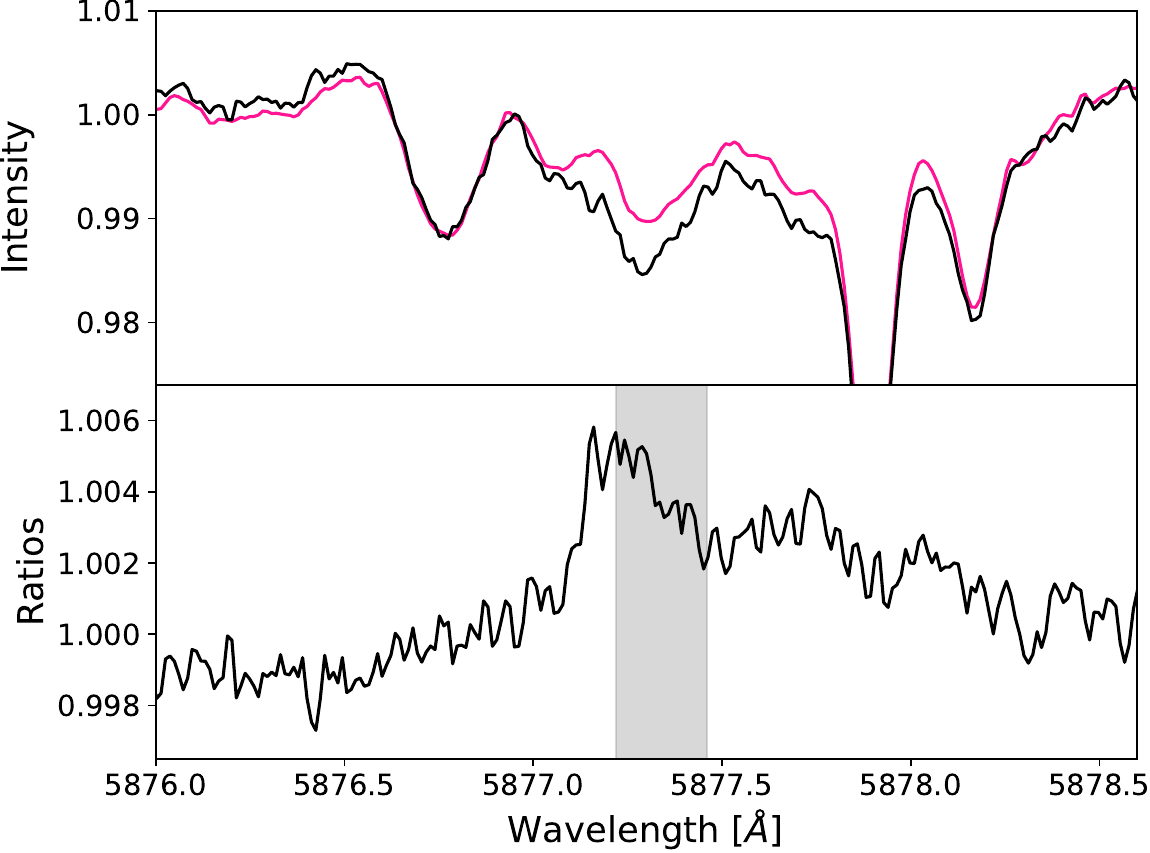}
    \caption{\textit{Top:} EXPRES master spectrum (pink), and one single observation (black) of the He I D3 line. \textit{Bottom:} Ratio of this wavelength range obtained by dividing the master spectrum by the single observation. The He D3 index is obtained by integrating the gray-shaded area under these ratios.}
    \label{fig:HeD3_resi}
\end{figure}

\section{Results}

\subsection{He D3 as Proxy for Magnetic Activity} \label{sec:HeD3}

While examining the variations of the He D3 triplet, we first found a general trend in the spectral region over time. In Fig.\,\ref{fig:HeD3_years}, we show yearly-averaged EXPRES spectra from the solar minimum towards the solar maximum. Notably, the region around 5877.3\,\AA\,(highlighted by the gray area) exhibits the most pronounced change in line depth over the years and gets deeper with the increasing solar activity.

The laboratory wavelengths of the Helium transition are shown in Table \ref{tab:table0}. The strongest He D3 transition is at $\lambda =$ 5877.24\,\AA\, \footnote{see https://www.chiantidatabase.org} indicating a wavelength shift of approximately 0.05\,\AA\,for our spectra. After EXPRES wavelength calibration, our precision reaches $10^{-4}\,$\AA\,, making it unlikely that this shift is due to instrumental uncertainty. Shallow lines can show an intrinsic red-shift \citep[e.g.][]{2016A&A...587A..65R, 2023A&A...680A..62E} but usually not stronger than a few hundreds of meters per second. Therefore, a shift of 0.05\,\AA\, seems too large since it would suggest a red-shift of about 2.5\,km\,s$^{-1}$. 
We therefore attribute the apparent offset to the presence of multiple He I triplet components, which overlap in this spectral region. However, since our goal is to use this line as an activity proxy, absolute wavelength calibration is not critical for our analysis.

\begin{table}[h]
\centering
\caption{He D3 Transitions}
\begin{tabular}{c c}
\hline
\hline
$\lambda$\,(vac)\,\AA & Transition\\
\hline
5877.2272 & 1s\,2p\,$^3P_2$ - 1s\,3d\,$^3D_1$\\
5877.2423 & 1s\,2p\,$^3P_2$ - 1s\,3d\,$^3D_2$\\
5877.2433 & 1s\,2p\,$^3P_2$ - 1s\,3d\,$^3D_3$\\
5877.2536 & 1s\,2p\,$^3P_1$ - 1s\,3d\,$^3D_1$\\
5877.2687 & 1s\,2p\,$^3P_1$ - 1s\,3d\,$^3D_2$\\
5877.5950 & 1s\,2p\,$^3P_0$ - 1s\,3d\,$^3D_1$\\
\hline
\end{tabular}

\label{tab:table0}
\vspace{2mm}
\begin{minipage}{0.9\linewidth}
\footnotesize \textit{Note.} Wavelength and transition information of the six possible \ch{He I} D3 transitions. Information taken from the NIST atomic spectra database (https://www.nist.gov/pml/atomic-spectra-database).
\end{minipage}

\end{table}

\begin{figure*}
    \centering
    \includegraphics[width=\linewidth]{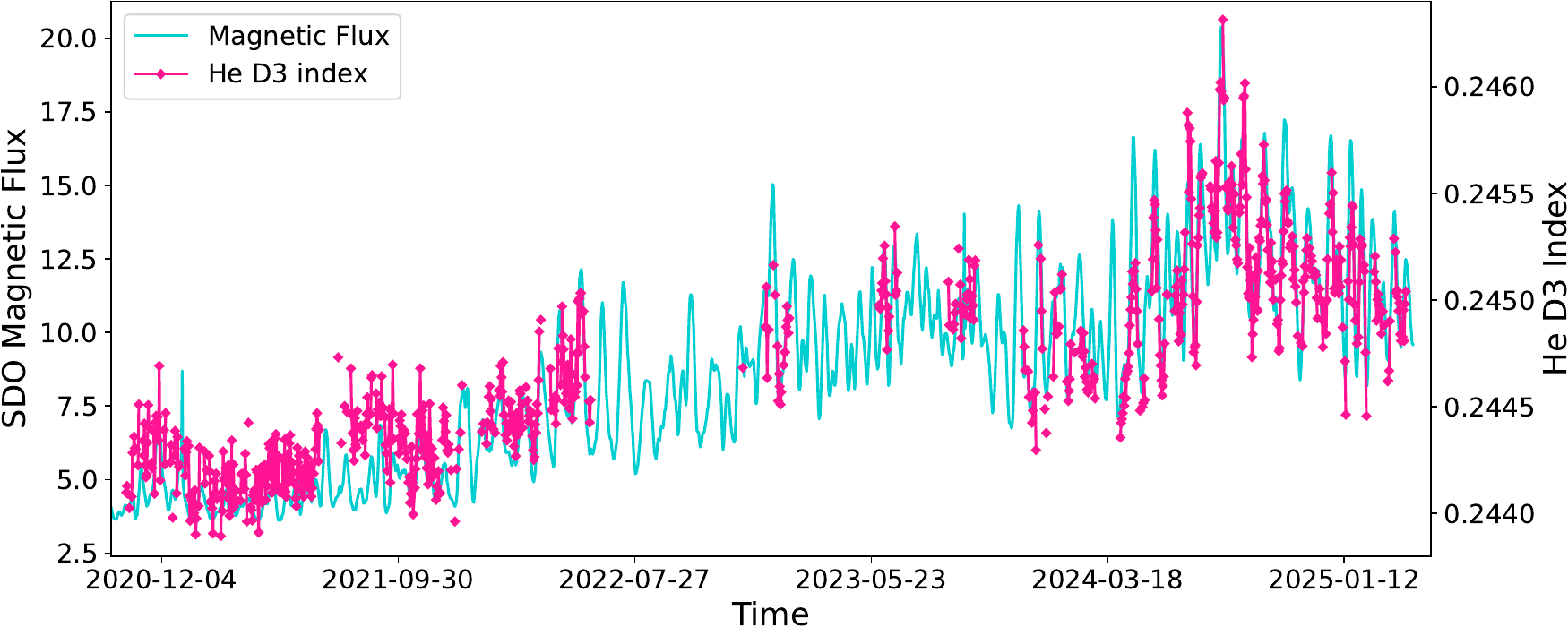}
    \caption{Magnetic activity indicators as a function of time. \textit{Cyan}: Unsigned magnetic flux calculated with \texttt{SolAster}. \textit{Magenta:} Daily \ch{He} D3 index derived from EXPRES solar observations. The long-term gap starting in 2022 was caused by a broken fiber in the setup, while the large observation gap in 2023 was due to smoke near the telescope site.}
    \label{fig:sdo_he0}
\end{figure*}

\begin{figure*}
    \centering
    \includegraphics[width=\linewidth]{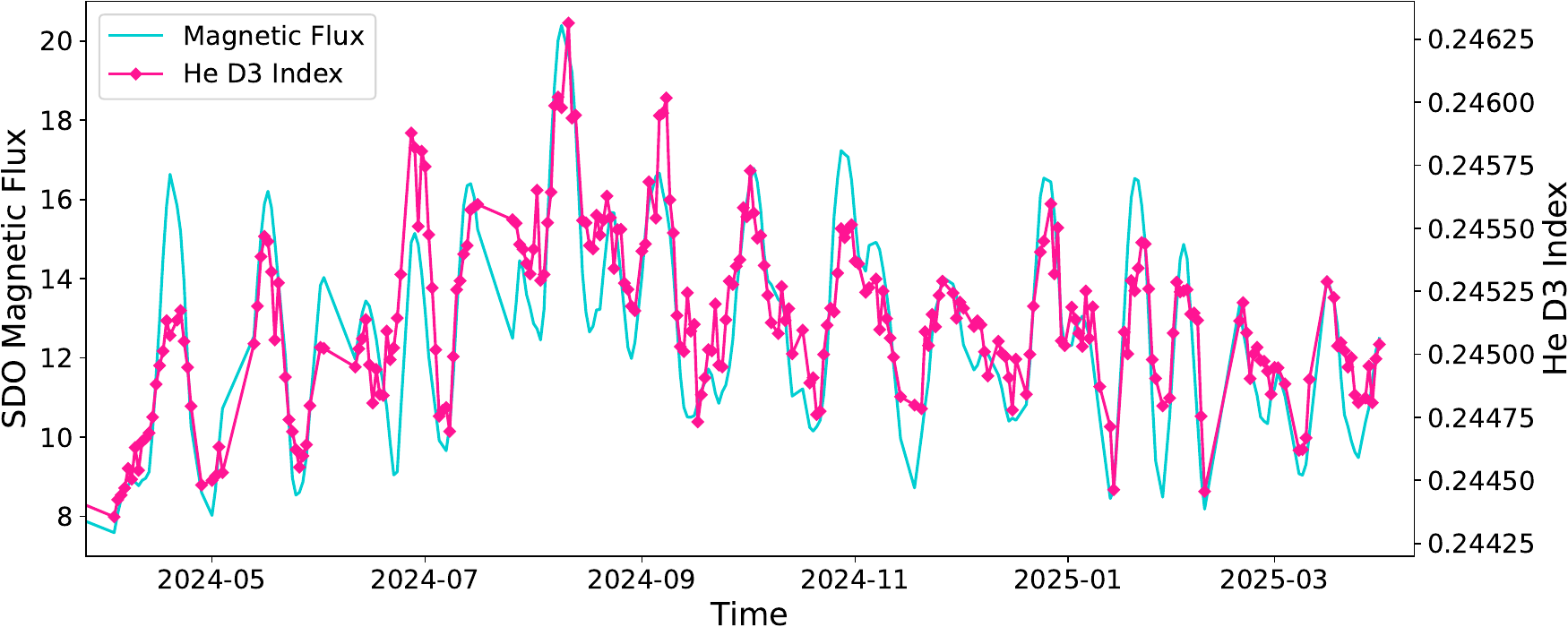}
    \caption{Same as Fig. \ref{fig:sdo_he0}, but zoomed in for the time range from April 2024 till end of March 2025 (during the solar maximum).}
    \label{fig:sdo_he}
\end{figure*}


To quantify the intensity variation of the He D3 line, we define a line index based on the residual flux, see Fig.\,\ref{fig:HeD3_resi}. The index is calculated by first constructing a master spectrum from all available EXPRES solar observations calculated from the median flux of each pixel. To achieve the best possible result, we also apply a linear fit to each single continuum in the wavelength region around the \ch{He} D3 line and shift the whole spectrum to the continuum level of the master spectrum. 
Each individual flux ratio was obtained by dividing the master spectrum by the corresponding daily spectrum. We then integrated the flux ratio within the target wavelength range $\lambda =$ (5877.22 - 5877.46)\,\AA\,(see gray area in Fig.\,\ref{fig:HeD3_resi}). This wavelength range was chosen since it coincides with the strongest line variance. We refer to this residual-based measurement as the He D3 index throughout the remainder of this paper.

Before calculating the He D3 index from the flux ratios, we first computed daily averaged spectra to increase the SNR and reduce short-term variability. The SNR for individual exposures in the relevant wavelength range is approximately 400, while the daily averaged spectra span a total SNR range from 400 to 1300, with a mean of 800. Individual He D3 index values derived from single spectra exhibit a typical intra-day standard deviation of $\sigma \approx 0.00015$ during solar minimum conditions, and up to $\sigma \approx 0.0003$ during times of increased activity around solar maximum.

In Fig.\ref{fig:sdo_he0}, the SDO unsigned magnetic flux and the He D3 index are shown as a function of time. To highlight the strongest activity phase, Fig.\ref{fig:sdo_he} provides a zoom in on the interval from April 2024 to the end of March 2025, during which the Sun reached its highest activity levels within our dataset. In this period of particularly pronounced magnetic variability, the He D3 index closely tracks the changes in the unsigned magnetic flux, demonstrating a robust correlation between the two quantities.

\begin{figure*}
    \centering
    \includegraphics[width=\linewidth]{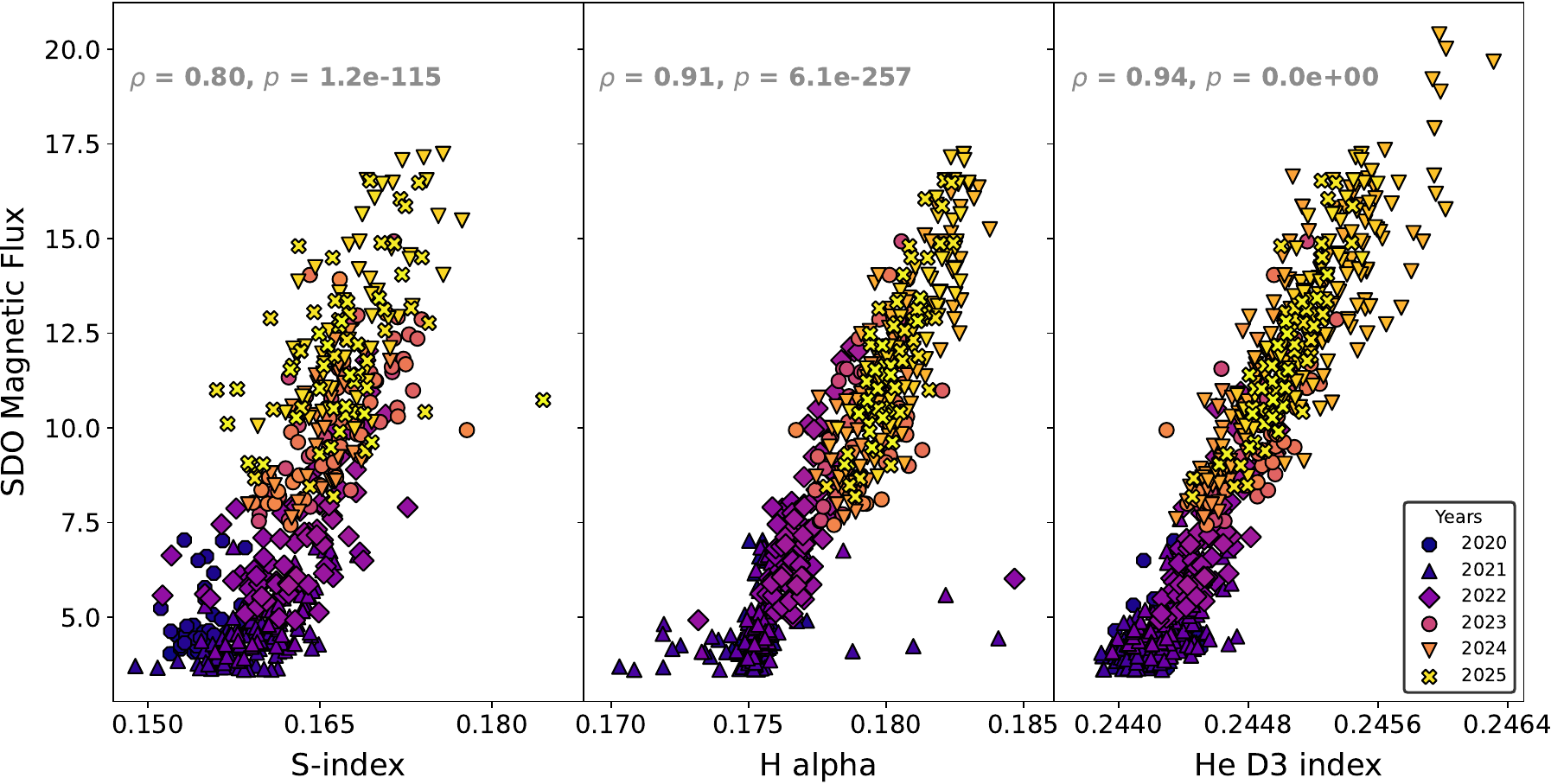}
    \caption{SDO unsigned magnetic flux plotted against the S-index, H alpha and the \ch{He} D3 index. Different symbols/colors denote data from individual years.}
    \label{fig:sdo_he_corr}
\end{figure*}
\begin{figure*}
    \centering
    \includegraphics[width=\linewidth]{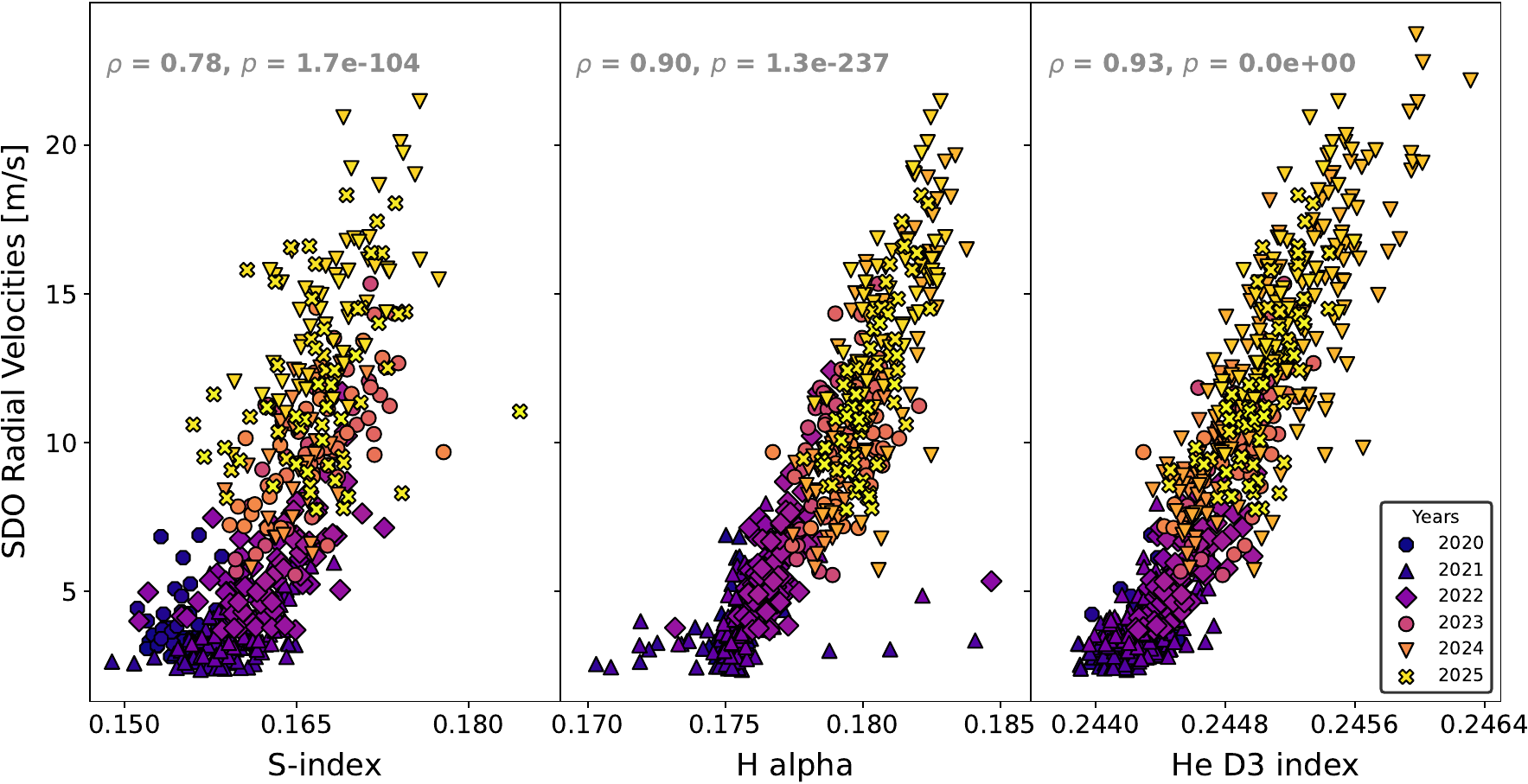}
    \caption{SDO RVs plotted against the S-index, H alpha and the the He D3 index. The different symbols/colors represent results of the different years.}
    \label{fig:sdo_he_corr2}
\end{figure*}

\subsection{Comparison with established Activity Indices}

To place the performance of the \ch{He I} D3 index into context, we compare it with established stellar activity indices, in particular, the widely used S-index derived from the Ca II H\&K lines and the H$\alpha$ index.

The S-index is one of the most widely used chromospheric activity indicators in stellar astrophysics \citep[e.g.][]{2000A&AS..142..275S,2010ApJ...725..875I}. It is based on the variation of strength of the \ch{Ca II} H and K emission lines at 3934.8\,\AA\, and 3969.6\,\AA\,(vacuum wavelength), respectively, which form in the chromosphere and are sensitive to magnetic activity. Originally developed in the context of the Mount Wilson Observatory project \citep{1978PASP...90..267V}, the S-index has since been applied extensively to study stellar activity cycles, rotation, and variability. On the other hand, the H$\alpha$ line at 6565\,\AA (vacuum wavelength) traces conditions in the middle chromosphere and also responds sensitively to plages, filaments, and other magnetic structures. Variations in H$\alpha$ absorption or emission strength have long been employed as tracers of magnetic activity in both the Sun and other cool stars \citep[e.g.][]{2002AJ....123.3356G, 2013ApJ...764....3R}. Taken together, the S-index and the H$\alpha$ index represent two of the most established diagnostics of stellar magnetic activity. They therefore provide a benchmark against which we can directly assess the performance of our \ch{He} D3 index.

We derive both activity indices from the same EXPRES solar observations that we use to calculate the \ch{He} D3 index. However, the S-index has some missing daily values where the calculation was not feasible (cf. Fig. \ref{fig:sdo_he_corr}) due to a low SNR in the respective wavelength area. Figure \ref{fig:sdo_he_corr} shows the unsigned magnetic flux from SDO against the S-index, the H$\alpha$ index, and the \ch{He} D3 index, respectively. In general, the years 2020 and 2021 do not show strong magnetic activity since this was the time during the solar minimum. However, all three values change in the same way as soon as the Sun became more active over time. We find a general Spearman correlation coefficient between the unsigned magnetic flux and the S-index of $\rho = 0.80$ while H\,$\alpha$ correlates with $\rho = 0.91$. In comparison, the correlation with the He D3 index is $\rho = 0.94$. This stronger correlation exhibited by the He D3 is not due to the additional data points that are missing in the S-index. Even when He D3 is restricted to the same dates as the ones which are available for the S-index, the correlation of 0.94 is preserved. We investigate the correlations between the unsigned magnetic flux and the He D3 index by year; the correlation strengthens with time as the Sun reaches its activity maximum, cf. Table\,\ref{tab:table1}.

\begin{table}[h]
\centering
\begin{tabular}{ccccccc}
\hline
\hline
Year & 2020 & 2021 & 2022 & 2023 & 2024 & 2025 \\
\hline
$\rho$ & 0.26 & 0.54 & 0.76 & 0.64 & 0.84 & 0.82 \\
$p$    & 0.12 & $10^{-20}$ & $10^{-19}$ & $10^{-10}$ & $10^{-67}$ & $10^{-16}$ \\
\hline
\end{tabular}
\caption{Spearman Correlation between the unsigned magnetic flux received by SDO and the He D3 index per year}
\label{tab:table1}
\end{table}

We also calculated the correlation between the three indices and the SDO RVs (Fig.\,\ref{fig:sdo_he_corr2}). A similar pattern emerges: the \ch{He} D3 index shows the strongest correlation with the RVs ($\rho = 0.93$) followed by H$\alpha$ ($\rho = 0.90$) and the S-index ($\rho = 0.78$). These results demonstrate that the \ch{He} D3 index performance is on par with established indicators such as H$\alpha$ and the S-index and may even outperform them under certain conditions. This underlines its potential as a valuable addition to the set of activity proxies.

\subsection{Differential Rotation}
Since stellar rotation often modulates chromospheric activity signals, we further investigated whether the temporal behavior of the \ch{He} D3 index reflects the Sun’s differential rotation. Detecting such signatures is particularly challenging as the variation in the solar rotation period across different latitudes is modest and can be masked by short-term variability and noise. To search for periodic behavior in the time series of the \ch{He} D3 index, we computed Lomb–Scargle periodograms \citep{1976Ap&SS..39..447L, 1982ApJ...263..835S}. Probing the Sun’s differential rotation with this approach is challenging since the variance of rotation period is quite small. Sunspots and associated magnetic regions typically emerge at heliographic latitudes between $\sim$35° and $\sim$5° (with 0° denoting the equator), where the rotation period changes from about 26.7 to 25.0 days over the course of the cycle \citep[cf.][]{2021SoPh..296...25J, 2025arXiv250808196L}. Accordingly, the measured rotation period is expected to decrease from $\sim$26.7 days in the early phase of the cycle (around 2020) to $\sim$25.0 days closer to maximum activity (around 2025) as spot emergence migrates from $\sim$30 degrees toward the equator as the cycle progresses.

\begin{figure}
    \centering
    \includegraphics[width=\linewidth]{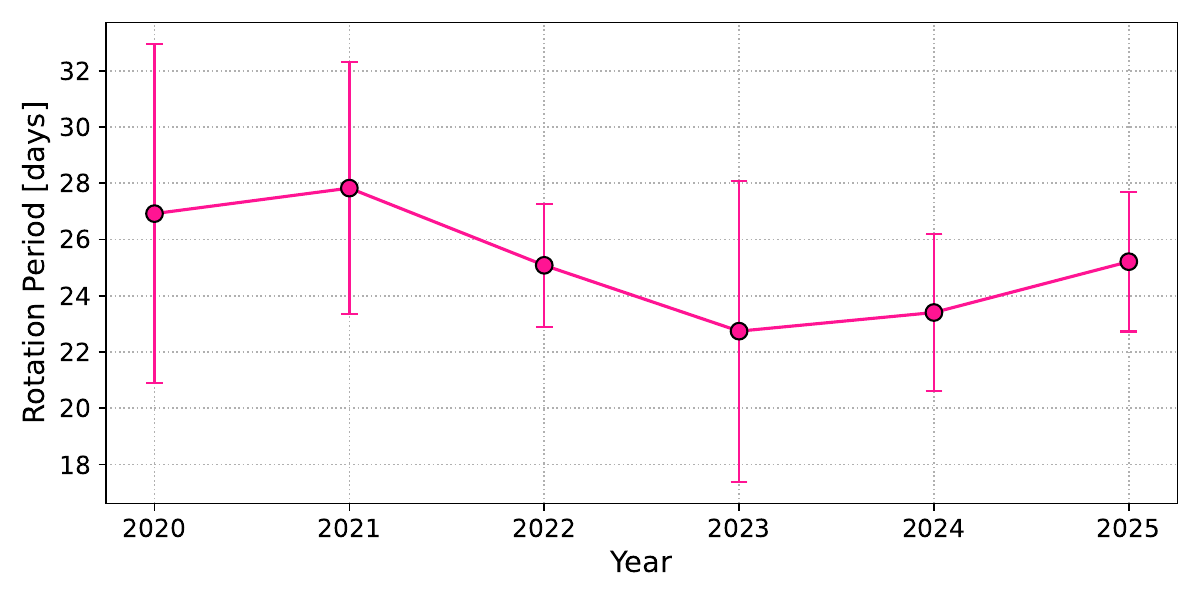}
    \caption{Obtained rotation period per year of observation using the Lomb-Scargle periodogram of the He D3 index time series.}
    \label{fig:rotation_periode}
\end{figure}

To further investigate this, we determined the rotation period separately for each year, restricting the period search to 10–100 days. The results, shown in Fig.\,\ref{fig:rotation_periode}, roughly follow the expected negative trend of differential rotation with larger uncertainties during periods of low activity (2020–2021) and sparser sampling (2023; cf. Fig.\,\ref{fig:sdo_he0}). Averaging the yearly estimates yields a weighted mean rotation period of $24.9 \pm 1.3$ days, which agrees with the known solar rotation period when the uncertainties are taken into account.
The slightly shorter total period we found compared to sunspot-based values may arise because the \ch{He} D3 index does not only trace sunspots but also responds to plages, filaments, and other chromospheric features. These structures may form closer to the equator where the solar rotation is faster \citep{1986A&A...155...87B}, thus biasing the signal toward shorter periods. 
Despite these subtleties, the consistency of the annual and mean periods demonstrates that the \ch{He} D3 index robustly traces solar rotation.

\section{Summary and Conclusion}

In this study, we revisited the magnetically sensitive \ch{He} D3 spectral line and analyzed its variability in the solar spectrum using EXPRES data. Our goal was to re-evaluate this line as a potential activity proxy. We show that the \ch{He} D3 index serves as a robust and reliable tracer of both the solar magnetic flux and RV variations with Spearman correlation coefficients of $\rho = 0.94$ and $\rho = 0.93$, respectively. These results highlight its potential as a valuable activity indicator for Sun-like stars. For comparison, the widely used S-index reaches lower correlations of only $\rho = 0.80$ with magnetic flux and $\rho = 0.78$ with RVs in our dataset while H$\alpha$ shows slightly higher values of $\rho = 0.91$ with magnetic flux and $\rho = 0.90$ with RVs.

To place our findings in a wider context, \citet{2022ApJ...935....6H} explored a variety of activity parameters in the solar case and assessed their correlation with solar RVs and the observed magnetic field. They reported a Spearman correlation coefficient of $\rho = 0.92$ between solar RVs and the unsigned magnetic flux, whereas the S-index exhibited a lower correlation with solar RVs of only $\rho = 0.75$. 

An advantage of the He D3 line is its location in the redder part of the spectrum ($\sim 5877$\,\AA\,compared to the Ca lines near $\sim 3935$\,\AA\,and $\sim 3970$\,\AA). This shift translates to a generally higher photon flux and improved sensitivity for modern CCD detectors, making it easier to detect line variations in the longer wavelength region. One potential concern is that equal variations in magnetic flux ($\Delta{\rm mag flux}\,\simeq\,10$) correspond to smaller changes in the He D3 index ($\Delta\,\text{HeD3}\,\approx\,$0.001) compared to the S-index ($\Delta\,$S-index$\,\approx$\,0.01) or H$\alpha$ ($\Delta\,\text{H}\,\alpha\,\approx$\,0.01) (see Fig. \ref{fig:sdo_he_corr}). However, despite this smaller absolute response, the He D3 index maintains a very strong correlation with both the magnetic flux and RVs, thereby overcoming this possible disadvantage and reinforcing its value as a robust activity proxy.

Additionally, we used the new He D3 index to derive the solar rotation period. We found an expected negative trend of the rotation period over time, as the observations covered the time from solar minimum to solar maximum. We determined a total solar rotation period of $24.9 \pm 1.3$ days.

While our analysis demonstrates that the \ch{He} D3 index can be robustly measured from daily averaged EXPRES solar spectra, its applicability to other stars depends on several observational factors. In typical RV surveys, spectra are often obtained at lower SNR levels ($\sim$250 per exposure), where the weak \ch{He} D3 line becomes more strongly affected by photon noise.
To test how sensitive the index is to noise, we repeated our correlation analysis using single (non-averaged) solar observations, which have an SNR of about 500 in the relevant wavelength range. In this case, the correlation with both unsigned magnetic flux and solar RVs weakened substantially, dropping to $\rho \approx 0.78$ and $\rho \approx 0.76$, respectively. This is in contrast to the daily averaged spectra used in the main analysis, which span a total SNR range of 400–1300, with a mean SNR of 800.
These findings suggest that while the He D3 index can still be measured at lower SNR, a higher SNR—preferably above $\sim$800—is beneficial to fully recover the strong correlations with other magnetic activity indicators.

Another important factor is spectral resolution: the high resolving power of EXPRES ($R \approx 137{,}500$) allows nearby blends to be separated more clearly, whereas at lower resolutions the reliability of this index may decrease. Finally, stellar rotation provides an additional limitation. For stars with higher projected rotational velocities ($v\sin i$), rotational broadening significantly widens spectral lines, diluting the \ch{He} D3 feature and making activity-driven variability harder to detect. Taken together, these considerations suggest that the \ch{He} D3 index is particularly promising for slowly rotating, Sun-like stars observed with high-resolution spectrographs at sufficiently high SNR.

Additionally, the applicability of the \ch{He I} D3 index may not extend universally across all stellar types. Studies of M dwarfs have shown that only a minority exhibit significant activity-induced variability in the \ch{He} D3 line \citep{2009MNRAS.400..238H, 2011A&A...534A..30G}. For instance, \citet{2011A&A...534A..30G} found that only about 10\% of their M dwarf sample displayed detectable variability in this line, leading them to conclude that \ch{He} D3 is not well suited for routine activity monitoring in M-type stars.

Taken together, our findings propose that the \ch{He} D3 index holds strong promise as an activity proxy for Sun-like stars. It may serve as a complementary or even superior activity indicator under certain conditions. Future studies extending this analysis to other spectral types and stellar inclinations will help further refine its utility in the broader context of stellar activity mitigation.

\begin{acknowledgments}
M.E. and J.L. acknowledge support from NASA under award No. 80NSSC23K0040. These results made use of the Lowell Observatory Solar Telescope that was generously funded through the Mt. Cuba Astronomical Foundation and the Lowell Discovery Telescope at Lowell Observatory. Lowell is a private, non-profit institution dedicated to astrophysical research and public appreciation of astronomy and operates the LDT in partnership with Boston University, the University of Maryland, the University of Toledo, Northern Arizona University and Yale University. Lowell Observatory sits at the base of mountains sacred to tribes throughout the region. We honor their past, present, and future generations, who have lived here for millennia and will forever call this place home. The EXPRES  team acknowledges support for the design and construction of EXPRES from NSF MRI-1429365, NSF ATI-1509436 and Yale University. DAF gratefully acknowledges support to carry out this research from NSF 2009528, NSF 1616086, NSF AST-2009528, the Heising-Simons Foundation, and an anonymous donor in the Yale alumni community. Additionally, we thank the anonymous referee for their helpful comments, which improved the clarity of this manuscript.
\end{acknowledgments}

\bibliography{bib}{}

\begin{thebibliography}{}
\expandafter\ifx\csname natexlab\endcsname\relax\def\natexlab#1{#1}\fi
\providecommand{\url}[1]{\href{#1}{#1}}
\providecommand{\dodoi}[1]{doi:~\href{http://doi.org/#1}{\nolinkurl{#1}}}
\providecommand{\doeprint}[1]{\href{http://ascl.net/#1}{\nolinkurl{http://ascl.net/#1}}}
\providecommand{\doarXiv}[1]{\href{https://arxiv.org/abs/#1}{\nolinkurl{https://arxiv.org/abs/#1}}}

\bibitem[{H. {Balthasar} {et~al.}(1986){Balthasar}, {Vazquez}, \& {Woehl}}]{1986A&A...155...87B}
{Balthasar}, H., {Vazquez}, M., \& {Woehl}, H. 1986, \bibinfo{title}{{Differential rotation of sunspot groups in the period from 1874 through 1976 and changes of the rotation velocity within the solar cycle},} \aap, 155, 87

\bibitem[{N.~E. {Batalha} {et~al.}(2019){Batalha}, {Lewis}, {Fortney}, {Batalha}, {Kempton}, {Lewis}, \& {Line}}]{2019ApJ...885L..25B}
{Batalha}, N.~E., {Lewis}, T., {Fortney}, J.~J., {et~al.} 2019, \bibinfo{title}{{The Precision of Mass Measurements Required for Robust Atmospheric Characterization of Transiting Exoplanets},} \apjl, 885, L25, \dodoi{10.3847/2041-8213/ab4909}

\bibitem[{W.~J. {Chaplin} {et~al.}(1996){Chaplin}, {Elsworth}, {Howe}, {Isaak}, {McLeod}, {Miller}, {van der Raay}, {Wheeler}, \& {New}}]{1996SoPh..168....1C}
{Chaplin}, W.~J., {Elsworth}, Y., {Howe}, R., {et~al.} 1996, \bibinfo{title}{{BiSON Performance},} \solphys, 168, 1, \dodoi{10.1007/BF00145821}

\bibitem[{R. {Cosentino} {et~al.}(2012){Cosentino}, {Lovis}, {Pepe}, {Collier Cameron}, {Latham}, {Molinari}, {Udry}, {Bezawada}, {Black}, {Born}, {Buchschacher}, {Charbonneau}, {Figueira}, {Fleury}, {Galli}, {Gallie}, {Gao}, {Ghedina}, {Gonzalez}, {Gonzalez}, {Guerra}, {Henry}, {Horne}, {Hughes}, {Kelly}, {Lodi}, {Lunney}, {Maire}, {Mayor}, {Micela}, {Ordway}, {Peacock}, {Phillips}, {Piotto}, {Pollacco}, {Queloz}, {Rice}, {Riverol}, {Riverol}, {San Juan}, {Sasselov}, {Segransan}, {Sozzetti}, {Sosnowska}, {Stobie}, {Szentgyorgyi}, {Vick}, \& {Weber}}]{2012SPIE.8446E..1VC}
{Cosentino}, R., {Lovis}, C., {Pepe}, F., {et~al.} 2012, in Society of Photo-Optical Instrumentation Engineers (SPIE) Conference Series, Vol. 8446, Ground-based and Airborne Instrumentation for Astronomy IV, ed. I.~S. {McLean}, S.~K. {Ramsay}, \& H.~{Takami}, 84461V, \dodoi{10.1117/12.925738}

\bibitem[{J. {Crass} {et~al.}(2021){Crass}, {Gaudi}, {Leifer}, {Beichman}, {Bender}, {Blackwood}, {Burt}, {Callas}, {Cegla}, {Diddams}, {Dumusque}, {Eastman}, {Ford}, {Fulton}, {Gibson}, {Halverson}, {Haywood}, {Hearty}, {Howard}, {Latham}, {L{\"o}hner-B{\"o}ttcher}, {Mamajek}, {Mortier}, {Newman}, {Plavchan}, {Quirrenbach}, {Reiners}, {Robertson}, {Roy}, {Schwab}, {Seifahrt}, {Szentgyorgyi}, {Terrien}, {Teske}, {Thompson}, \& {Vasisht}}]{2021arXiv210714291C}
{Crass}, J., {Gaudi}, B.~S., {Leifer}, S., {et~al.} 2021, \bibinfo{title}{{Extreme Precision Radial Velocity Working Group Final Report},} arXiv e-prints, arXiv:2107.14291, \dodoi{10.48550/arXiv.2107.14291}

\bibitem[{X. {Dumusque} {et~al.}(2011){Dumusque}, {Santos}, {Udry}, {Lovis}, \& {Bonfils}}]{2011A&A...527A..82D}
{Dumusque}, X., {Santos}, N.~C., {Udry}, S., {Lovis}, C., \& {Bonfils}, X. 2011, \bibinfo{title}{{Planetary detection limits taking into account stellar noise. II. Effect of stellar spot groups on radial-velocities},} \aap, 527, A82, \dodoi{10.1051/0004-6361/201015877}

\bibitem[{X. {Dumusque} {et~al.}(2015){Dumusque}, {Glenday}, {Phillips}, {Buchschacher}, {Collier Cameron}, {Cecconi}, {Charbonneau}, {Cosentino}, {Ghedina}, {Latham}, {Li}, {Lodi}, {Lovis}, {Molinari}, {Pepe}, {Udry}, {Sasselov}, {Szentgyorgyi}, \& {Walsworth}}]{2015ApJ...814L..21D}
{Dumusque}, X., {Glenday}, A., {Phillips}, D.~F., {et~al.} 2015, \bibinfo{title}{{HARPS-N Observes the Sun as a Star},} \apjl, 814, L21, \dodoi{10.1088/2041-8205/814/2/L21}

\bibitem[{M. {Ellwarth} {et~al.}(2023){Ellwarth}, {Ehmann}, {Sch{\"a}fer}, \& {Reiners}}]{2023A&A...680A..62E}
{Ellwarth}, M., {Ehmann}, B., {Sch{\"a}fer}, S., \& {Reiners}, A. 2023, \bibinfo{title}{{Convective characteristics of Fe I lines across the solar disc},} \aap, 680, A62, \dodoi{10.1051/0004-6361/202347615}

\bibitem[{T. {Ervin} {et~al.}(2022){Ervin}, {Halverson}, {Burrows}, {Murphy}, {Roy}, {Haywood}, {Rescigno}, {Bender}, {Lin}, {Burt}, \& {Mahadevan}}]{2022AJ....163..272E}
{Ervin}, T., {Halverson}, S., {Burrows}, A., {et~al.} 2022, \bibinfo{title}{{Leveraging Space-based Data from the Nearest Solar-type Star to Better Understand Stellar Activity Signatures in Radial Velocity Data},} \aj, 163, 272, \dodoi{10.3847/1538-3881/ac67e6}

\bibitem[{S.~R. {Gibson} {et~al.}(2016){Gibson}, {Howard}, {Marcy}, {Edelstein}, {Wishnow}, \& {Poppett}}]{2016SPIE.9908E..70G}
{Gibson}, S.~R., {Howard}, A.~W., {Marcy}, G.~W., {et~al.} 2016, in Society of Photo-Optical Instrumentation Engineers (SPIE) Conference Series, Vol. 9908, Ground-based and Airborne Instrumentation for Astronomy VI, ed. C.~J. {Evans}, L.~{Simard}, \& H.~{Takami}, 990870, \dodoi{10.1117/12.2233334}

\bibitem[{J.~E. {Gizis} {et~al.}(2002){Gizis}, {Reid}, \& {Hawley}}]{2002AJ....123.3356G}
{Gizis}, J.~E., {Reid}, I.~N., \& {Hawley}, S.~L. 2002, \bibinfo{title}{{The Palomar/MSU Nearby Star Spectroscopic Survey. III. Chromospheric Activity, M Dwarf Ages, and the Local Star Formation History},} \aj, 123, 3356, \dodoi{10.1086/340465}

\bibitem[{J. {Gomes da Silva} {et~al.}(2011){Gomes da Silva}, {Santos}, {Bonfils}, {Delfosse}, {Forveille}, \& {Udry}}]{2011A&A...534A..30G}
{Gomes da Silva}, J., {Santos}, N.~C., {Bonfils}, X., {et~al.} 2011, \bibinfo{title}{{Long-term magnetic activity of a sample of M-dwarf stars from the HARPS program. I. Comparison of activity indices},} \aap, 534, A30, \dodoi{10.1051/0004-6361/201116971}

\bibitem[{J.~W. {Harvey} {et~al.}(1996){Harvey}, {Hill}, {Hubbard}, {Kennedy}, {Leibacher}, {Pintar}, {Gilman}, {Noyes}, {Title}, {Toomre}, {Ulrich}, {Bhatnagar}, {Kennewell}, {Marquette}, {Patron}, {Saa}, \& {Yasukawa}}]{1996Sci...272.1284H}
{Harvey}, J.~W., {Hill}, F., {Hubbard}, R.~P., {et~al.} 1996, \bibinfo{title}{{The Global Oscillation Network Group (GONG) Project},} Science, 272, 1284, \dodoi{10.1126/science.272.5266.1284}

\bibitem[{R.~D. {Haywood} {et~al.}(2016){Haywood}, {Collier Cameron}, {Unruh}, {Lovis}, {Lanza}, {Llama}, {Deleuil}, {Fares}, {Gillon}, {Moutou}, {Pepe}, {Pollacco}, {Queloz}, \& {S{\'e}gransan}}]{2016MNRAS.457.3637H}
{Haywood}, R.~D., {Collier Cameron}, A., {Unruh}, Y.~C., {et~al.} 2016, \bibinfo{title}{{The Sun as a planet-host star: proxies from SDO images for HARPS radial-velocity variations},} \mnras, 457, 3637, \dodoi{10.1093/mnras/stw187}

\bibitem[{R.~D. {Haywood} {et~al.}(2022){Haywood}, {Milbourne}, {Saar}, {Mortier}, {Phillips}, {Charbonneau}, {Cameron}, {Cegla}, {Meunier}, \& {}}]{2022ApJ...935....6H}
{Haywood}, R.~D., {Milbourne}, T.~W., {Saar}, S.~H., {et~al.} 2022, \bibinfo{title}{{Unsigned Magnetic Flux as a Proxy for Radial-velocity Variations in Sun-like Stars},} \apj, 935, 6, \dodoi{10.3847/1538-4357/ac7c12}

\bibitem[{E.~R. {Houdebine} {et~al.}(2009){Houdebine}, {Stempels}, \& {Oliveira}}]{2009MNRAS.400..238H}
{Houdebine}, E.~R., {Stempels}, H.~C., \& {Oliveira}, J.~H. 2009, \bibinfo{title}{{Observation and modelling of main sequence star chromospheres - XIII. The NaI D1 and D2, and HeI D3 lines in dM1 stars},} \mnras, 400, 238, \dodoi{10.1111/j.1365-2966.2009.15433.x}

\bibitem[{H. {Isaacson} \& D. {Fischer}(2010){Isaacson} \& {Fischer}}]{2010ApJ...725..875I}
{Isaacson}, H., \& {Fischer}, D. 2010, \bibinfo{title}{{Chromospheric Activity and Jitter Measurements for 2630 Stars on the California Planet Search},} \apj, 725, 875, \dodoi{10.1088/0004-637X/725/1/875}

\bibitem[{B.~K. {Jha} {et~al.}(2021){Jha}, {Priyadarshi}, {Mandal}, {Chatterjee}, \& {Banerjee}}]{2021SoPh..296...25J}
{Jha}, B.~K., {Priyadarshi}, A., {Mandal}, S., {Chatterjee}, S., \& {Banerjee}, D. 2021, \bibinfo{title}{{Measurements of Solar Differential Rotation Using the Century Long Kodaikanal Sunspot Data},} \solphys, 296, 25, \dodoi{10.1007/s11207-021-01767-8}

\bibitem[{C. {Jurgenson} {et~al.}(2016){Jurgenson}, {Fischer}, {McCracken}, {Sawyer}, {Szymkowiak}, {Davis}, {Muller}, \& {Santoro}}]{2016SPIE.9908E..6TJ}
{Jurgenson}, C., {Fischer}, D., {McCracken}, T., {et~al.} 2016, in Society of Photo-Optical Instrumentation Engineers (SPIE) Conference Series, Vol. 9908, Ground-based and Airborne Instrumentation for Astronomy VI, ed. C.~J. {Evans}, L.~{Simard}, \& H.~{Takami}, 99086T, \dodoi{10.1117/12.2233002}

\bibitem[{C.~U. {Keller} \&  {Nso Staff}(1998){Keller} \& {Nso Staff}}]{1998ASPC..154..636K}
{Keller}, C.~U., \& {Nso Staff}. 1998, in Astronomical Society of the Pacific Conference Series, Vol. 154, Cool Stars, Stellar Systems, and the Sun, ed. R.~A. {Donahue} \& J.~A. {Bookbinder}, 636

\bibitem[{B. {Klein} {et~al.}(2022){Klein}, {Zicher}, {Kavanagh}, {Nielsen}, {Aigrain}, {Vidotto}, {Barrag{\'a}n}, {Strugarek}, {Nicholson}, {Donati}, \& {Bouvier}}]{2022MNRAS.512.5067K}
{Klein}, B., {Zicher}, N., {Kavanagh}, R.~D., {et~al.} 2022, \bibinfo{title}{{One year of AU Mic with HARPS - II. Stellar activity and star-planet interaction},} \mnras, 512, 5067, \dodoi{10.1093/mnras/stac761}

\bibitem[{D.~A. {Landman}(1981){Landman}}]{1981ApJ...244..345L}
{Landman}, D.~A. 1981, \bibinfo{title}{{Measurements of He D3 profiles in solar plages},} \apj, 244, 345, \dodoi{10.1086/158712}

\bibitem[{C. {Leet} {et~al.}(2019){Leet}, {Fischer}, \& {Valenti}}]{leet2019}
{Leet}, C., {Fischer}, D.~A., \& {Valenti}, J.~A. 2019, \bibinfo{title}{{Toward a Self-calibrating, Empirical, Light-weight Model for Tellurics in High-resolution Spectra},} \aj, 157, 187, \dodoi{10.3847/1538-3881/ab0d86}

\bibitem[{S.~E. {Levine} {et~al.}(2012){Levine}, {Bida}, {Chylek}, {Collins}, {DeGroff}, {Dunham}, {Lotz}, {Venetiou}, \& {Zoonemat Kermani}}]{2012SPIE.8444E..19L}
{Levine}, S.~E., {Bida}, T.~A., {Chylek}, T., {et~al.} 2012, in Society of Photo-Optical Instrumentation Engineers (SPIE) Conference Series, Vol. 8444, Ground-based and Airborne Telescopes IV, ed. L.~M. {Stepp}, R.~{Gilmozzi}, \& H.~J. {Hall}, 844419, \dodoi{10.1117/12.926415}

\bibitem[{A.~S.~J. {Lin} {et~al.}(2022){Lin}, {Monson}, {Mahadevan}, {Ninan}, {Halverson}, {Nitroy}, {Bender}, {Logsdon}, {Kanodia}, {Terrien}, {Roy}, {Luhn}, {Gupta}, {Ford}, {Hearty}, {Laher}, {Hunting}, {McBride}, {Salazar Rivera}, {Rajagopal}, {Wolf}, {Robertson}, {Wright}, {Blake}, {Ca{\~n}as}, {Lubar}, {McElwain}, {Ramsey}, {Schwab}, \& {Stefansson}}]{2022AJ....163..184L}
{Lin}, A. S.~J., {Monson}, A., {Mahadevan}, S., {et~al.} 2022, \bibinfo{title}{{Observing the Sun as a Star: Design and Early Results from the NEID Solar Feed},} \aj, 163, 184, \dodoi{10.3847/1538-3881/ac5622}

\bibitem[{J. {Llama} {et~al.}(2022){Llama}, {Fischer}, {Brewer}, {Zhao}, \& {Szymkowiak}}]{2022BAAS...54e.102L}
{Llama}, J., {Fischer}, D., {Brewer}, J., {Zhao}, L., \& {Szymkowiak}, A. 2022, in Bulletin of the American Astronomical Society, Vol.~54, 102.102

\bibitem[{N.~R. {Lomb}(1976){Lomb}}]{1976Ap&SS..39..447L}
{Lomb}, N.~R. 1976, \bibinfo{title}{{Least-Squares Frequency Analysis of Unequally Spaced Data},} \apss, 39, 447, \dodoi{10.1007/BF00648343}

\bibitem[{E.~J. {L{\"o}{\ss}nitz} {et~al.}(2025){L{\"o}{\ss}nitz}, {Pietrow}, {Chakraborty}, {Verma}, {Kontogiannis}, {Balthasar}, {Denker}, \& {Lendl}}]{2025arXiv250808196L}
{L{\"o}{\ss}nitz}, E.~J., {Pietrow}, A. G.~M., {Chakraborty}, H., {et~al.} 2025, \bibinfo{title}{{Differential rotation of solar {\ensuremath{\alpha}}-sunspots and implications for stellar light curves},} arXiv e-prints, arXiv:2508.08196, \dodoi{10.48550/arXiv.2508.08196}

\bibitem[{N. {Meunier} {et~al.}(2010){Meunier}, {Desort}, \& {Lagrange}}]{2010A&A...512A..39M}
{Meunier}, N., {Desort}, M., \& {Lagrange}, A.~M. 2010, \bibinfo{title}{{Using the Sun to estimate Earth-like planets detection capabilities . II. Impact of plages},} \aap, 512, A39, \dodoi{10.1051/0004-6361/200913551}

\bibitem[{N. {Meunier} {et~al.}(2015){Meunier}, {Lagrange}, {Borgniet}, \& {Rieutord}}]{2015A&A...583A.118M}
{Meunier}, N., {Lagrange}, A.~M., {Borgniet}, S., \& {Rieutord}, M. 2015, \bibinfo{title}{{Using the Sun to estimate Earth-like planet detection capabilities. VI. Simulation of granulation and supergranulation radial velocity and photometric time series},} \aap, 583, A118, \dodoi{10.1051/0004-6361/201525721}

\bibitem[{J.~D. {Monnier} {et~al.}(2007){Monnier}, {Zhao}, {Pedretti}, {Thureau}, {Ireland}, {Muirhead}, {Berger}, {Millan-Gabet}, {Van Belle}, {ten Brummelaar}, {McAlister}, {Ridgway}, {Turner}, {Sturmann}, {Sturmann}, \& {Berger}}]{2007Sci...317..342M}
{Monnier}, J.~D., {Zhao}, M., {Pedretti}, E., {et~al.} 2007, \bibinfo{title}{{Imaging the Surface of Altair},} Science, 317, 342, \dodoi{10.1126/science.1143205}

\bibitem[{W.~D. {Pesnell} {et~al.}(2012){Pesnell}, {Thompson}, \& {Chamberlin}}]{2012SoPh..275....3P}
{Pesnell}, W.~D., {Thompson}, B.~J., \& {Chamberlin}, P.~C. 2012, \bibinfo{title}{{The Solar Dynamics Observatory (SDO)},} \solphys, 275, 3, \dodoi{10.1007/s11207-011-9841-3}

\bibitem[{R.~R. {Petersburg} {et~al.}(2020){Petersburg}, {Ong}, {Zhao}, {Blackman}, {Brewer}, {Buchhave}, {Cabot}, {Davis}, {Jurgenson}, {Leet}, {McCracken}, {Sawyer}, {Sharov}, {Tronsgaard}, {Szymkowiak}, \& {Fischer}}]{petersburg2020}
{Petersburg}, R.~R., {Ong}, J.~M.~J., {Zhao}, L.~L., {et~al.} 2020, \bibinfo{title}{{An Extreme-precision Radial-velocity Pipeline: First Radial Velocities from EXPRES},} \aj, 159, 187, \dodoi{10.3847/1538-3881/ab7e31}

\bibitem[{A. {Reiners} {et~al.}(2016){Reiners}, {Mrotzek}, {Lemke}, {Hinrichs}, \& {Reinsch}}]{2016A&A...587A..65R}
{Reiners}, A., {Mrotzek}, N., {Lemke}, U., {Hinrichs}, J., \& {Reinsch}, K. 2016, \bibinfo{title}{{The IAG solar flux atlas: Accurate wavelengths and absolute convective blueshift in standard solar spectra},} \aap, 587, A65, \dodoi{10.1051/0004-6361/201527530}

\bibitem[{P. {Robertson} {et~al.}(2013){Robertson}, {Endl}, {Cochran}, \& {Dodson-Robinson}}]{2013ApJ...764....3R}
{Robertson}, P., {Endl}, M., {Cochran}, W.~D., \& {Dodson-Robinson}, S.~E. 2013, \bibinfo{title}{{H{\ensuremath{\alpha}} Activity of Old M Dwarfs: Stellar Cycles and Mean Activity Levels for 93 Low-mass Stars in the Solar Neighborhood},} \apj, 764, 3, \dodoi{10.1088/0004-637X/764/1/3}

\bibitem[{R.~M. {Roettenbacher} {et~al.}(2016){Roettenbacher}, {Monnier}, {Korhonen}, {Aarnio}, {Baron}, {Che}, {Harmon}, {K{\H{o}}v{\'a}ri}, {Kraus}, {Schaefer}, {Torres}, {Zhao}, {Ten Brummelaar}, {Sturmann}, \& {Sturmann}}]{2016Natur.533..217R}
{Roettenbacher}, R.~M., {Monnier}, J.~D., {Korhonen}, H., {et~al.} 2016, \bibinfo{title}{{No Sun-like dynamo on the active star {\ensuremath{\zeta}} Andromedae from starspot asymmetry},} \nat, 533, 217, \dodoi{10.1038/nature17444}

\bibitem[{R.~A. {Rubenzahl} {et~al.}(2023){Rubenzahl}, {Halverson}, {Walawender}, {Hill}, {Howard}, {Brown}, {Ida}, {Tehero}, {Fulton}, {Gibson}, {Kassis}, {Smith}, {Wold}, \& {Payne}}]{2023PASP..135l5002R}
{Rubenzahl}, R.~A., {Halverson}, S., {Walawender}, J., {et~al.} 2023, \bibinfo{title}{{Staring at the Sun with the Keck Planet Finder: An Autonomous Solar Calibrator for High Signal-to-noise Sun-as-a-star Spectra},} \pasp, 135, 125002, \dodoi{10.1088/1538-3873/ad0b30}

\bibitem[{S.~H. {Saar} \& R.~A. {Donahue}(1997){Saar} \& {Donahue}}]{1997ApJ...485..319S}
{Saar}, S.~H., \& {Donahue}, R.~A. 1997, \bibinfo{title}{{Activity-Related Radial Velocity Variation in Cool Stars},} \apj, 485, 319, \dodoi{10.1086/304392}

\bibitem[{S.~H. {Saar} {et~al.}(1997){Saar}, {Huovelin}, {Osten}, \& {Shcherbakov}}]{1997A&A...326..741S}
{Saar}, S.~H., {Huovelin}, J., {Osten}, R.~A., \& {Shcherbakov}, A.~G. 1997, \bibinfo{title}{{HeI D3 absorption and its relation to rotation and activity in G and K dwarfs.},} \aap, 326, 741

\bibitem[{J.~D. {Scargle}(1982){Scargle}}]{1982ApJ...263..835S}
{Scargle}, J.~D. 1982, \bibinfo{title}{{Studies in astronomical time series analysis. II. Statistical aspects of spectral analysis of unevenly spaced data.},} \apj, 263, 835, \dodoi{10.1086/160554}

\bibitem[{P.~H. {Scherrer} {et~al.}(2012){Scherrer}, {Schou}, {Bush}, {Kosovichev}, {Bogart}, {Hoeksema}, {Liu}, {Duvall}, {Zhao}, {Title}, {Schrijver}, {Tarbell}, \& {Tomczyk}}]{2012SoPh..275..207S}
{Scherrer}, P.~H., {Schou}, J., {Bush}, R.~I., {et~al.} 2012, \bibinfo{title}{{The Helioseismic and Magnetic Imager (HMI) Investigation for the Solar Dynamics Observatory (SDO)},} \solphys, 275, 207, \dodoi{10.1007/s11207-011-9834-2}

\bibitem[{P. {Sch{\"o}fer} {et~al.}(2019){Sch{\"o}fer}, {Jeffers}, {Reiners}, {Shulyak}, {Fuhrmeister}, {Johnson}, {Zechmeister}, {Ribas}, {Quirrenbach}, {Amado}, {Caballero}, {Anglada-Escud{\'e}}, {Bauer}, {B{\'e}jar}, {Cort{\'e}s-Contreras}, {Dreizler}, {Guenther}, {Kaminski}, {K{\"u}rster}, {Lafarga}, {Montes}, {Morales}, {Pedraz}, \& {Tal-Or}}]{2019A&A...623A..44S}
{Sch{\"o}fer}, P., {Jeffers}, S.~V., {Reiners}, A., {et~al.} 2019, \bibinfo{title}{{The CARMENES search for exoplanets around M dwarfs. Activity indicators at visible and near-infrared wavelengths},} \aap, 623, A44, \dodoi{10.1051/0004-6361/201834114}

\bibitem[{C.~J. {Schrijver} \& C. {Zwaan}(2000){Schrijver} \& {Zwaan}}]{2000ssma.book.....S}
{Schrijver}, C.~J., \& {Zwaan}, C. 2000, {Solar and Stellar Magnetic Activity}

\bibitem[{C. {Schwab} {et~al.}(2016){Schwab}, {Rakich}, {Gong}, {Mahadevan}, {Halverson}, {Roy}, {Terrien}, {Robertson}, {Hearty}, {Levi}, {Monson}, {Wright}, {McElwain}, {Bender}, {Blake}, {St{\"u}rmer}, {Gurevich}, {Chakraborty}, \& {Ramsey}}]{2016SPIE.9908E..7HS}
{Schwab}, C., {Rakich}, A., {Gong}, Q., {et~al.} 2016, in Society of Photo-Optical Instrumentation Engineers (SPIE) Conference Series, Vol. 9908, Ground-based and Airborne Instrumentation for Astronomy VI, ed. C.~J. {Evans}, L.~{Simard}, \& H.~{Takami}, 99087H, \dodoi{10.1117/12.2234411}

\bibitem[{E. {Shkolnik} {et~al.}(2008){Shkolnik}, {Bohlender}, {Walker}, \& {Collier Cameron}}]{2008ApJ...676..628S}
{Shkolnik}, E., {Bohlender}, D.~A., {Walker}, G. A.~H., \& {Collier Cameron}, A. 2008, \bibinfo{title}{{The On/Off Nature of Star-Planet Interactions},} \apj, 676, 628, \dodoi{10.1086/527351}

\bibitem[{K. {Strassmeier} {et~al.}(2000){Strassmeier}, {Washuettl}, {Granzer}, {Scheck}, \& {Weber}}]{2000A&AS..142..275S}
{Strassmeier}, K., {Washuettl}, A., {Granzer}, T., {Scheck}, M., \& {Weber}, M. 2000, \bibinfo{title}{{The Vienna-KPNO search for Doppler-imaging candidate stars. I. A catalog of stellar-activity indicators for 1058 late-type Hipparcos stars},} \aaps, 142, 275, \dodoi{10.1051/aas:2000328}

\bibitem[{J.~R. {Varsik} {et~al.}(1984){Varsik}, {Wolff}, \& {Heasley}}]{1984BAAS...16..940V}
{Varsik}, J.~R., {Wolff}, S.~C., \& {Heasley}, J.~N. 1984, in Bulletin of the American Astronomical Society, Vol.~16, 940

\bibitem[{A.~H. {Vaughan} {et~al.}(1978){Vaughan}, {Preston}, \& {Wilson}}]{1978PASP...90..267V}
{Vaughan}, A.~H., {Preston}, G.~W., \& {Wilson}, O.~C. 1978, \bibinfo{title}{{Flux measurements of Ca II and K emission.},} \pasp, 90, 267, \dodoi{10.1086/130324}

\bibitem[{T.~N. {Woods} {et~al.}(2000){Woods}, {Rottman}, {Harder}, {Lawrence}, {McClintock}, {Kopp}, \& {Pankratz}}]{2000SPIE.4135..192W}
{Woods}, T.~N., {Rottman}, G.~J., {Harder}, J.~W., {et~al.} 2000, in Society of Photo-Optical Instrumentation Engineers (SPIE) Conference Series, Vol. 4135, Earth Observing Systems V, ed. W.~L. {Barnes}, 192--203, \dodoi{10.1117/12.494229}

\end{thebibliography}
\bibliographystyle{aasjournalv7}

\end{document}